# Long range antiferromagnetic order in a rocksalt high entropy oxide

Junjie Zhang,[†,*] Jiaqiang Yan,[†] S. Calder,[‡] Qiang Zheng,[†] Michael A. McGuire,[†] D. L. Abernathy,[‡] Yang Ren,[∥] Saul H. Lapidus,[∥] Katharine Page,[‡] Hong Zheng,[⊥] J. W. Freeland,[∥] John D. Budai,[†] R. P. Hermann[†,**]

[†]Materials Science and Technology Division, Oak Ridge National Laboratory, Oak Ridge, Tennessee 37831, USA.

[‡]Neutron Scattering Division, Oak Ridge National Laboratory, Oak Ridge, Tennessee 37831, USA.

[∥]X-ray Science Division, Advanced Photon Source, Argonne National Laboratory, Argonne, Illinois 60439, USA.

[⊥]Materials Science Division, Argonne National Laboratory, Argonne, Illinois 60439, USA

**Abstract:** We report for the first time the magnetic structure of the high entropy oxide $(Mg_{0.2}Co_{0.2}Ni_{0.2}Cu_{0.2}Zn_{0.2})O$ using neutron powder diffraction. This material exhibits a sluggish magnetic transition but possesses a long-range ordered antiferromagnetic ground state, as revealed by DC and AC magnetic susceptibility, elastic and inelastic neutron scattering measurements. The magnetic propagation wavevector is $k=(½, ½, ½)$ based on the cubic structure $Fm$-$3m$, and the magnetic structure consists of ferromagnetic sheets in the (111) planes with spins antiparallel between two neighboring planes. Inelastic neutron scattering reveals strong magnetic excitations at 100 K that survive up to room temperature. This work demonstrates that entropy-stabilized oxides represent a unique platform to study long range magnetic order with extreme chemical disorder.

High entropy oxides (HEOs), a class of materials based on a concept parallel to high entropy alloys,[1-2] are solid solutions stabilized by high configuration entropies. Recently, these materials have attracted much attention due to fundamental issues related to phase formation and materials design as well as their potential multifunctional properties that can emerge from their high configurational entropy and spin order.[3-7]

($Mg_{0.2}Co_{0.2}Ni_{0.2}Cu_{0.2}Zn_{0.2}$)O (hereafter MgO-HEO), the first entropy-stabilized oxide, was reported by Rost et al in 2015.[3] Their choice of this system was based on the diversity in structures, coordination and cationic radii of MgO (*Fm-3m*), CoO (*Fm-3m*), NiO (*Fm-3m*), CuO (*C2/c*) and ZnO (*P6₃mc*). This material and its substitution by aliovalent elements were reported to exhibit colossal dielectric constants[8] and superionic conductivity,[9] making this new family of materials very promising for applications such as large-$k$ dielectric materials and solid state electrolytes in solid-state batteries. Inspired by the superionic conductivity, Sarkar et al reported the electrochemical properties of the rocksalt high entropy oxides, such as lithium storage capacity and the cycling stability.[10] One of the fundamental interests in MgO-HEO is the possible Jahn-Teller distortion of $Cu^{2+}$ environment. Electron paramagnetic resonance,[11] extended x-ray absorption fine structure[12] and x-ray diffraction studies[11] as well as theoretical calculations[13-14] suggest this distortion. Besides studies of bulk samples, fundamental studies have been carried out on thin films.[15-16] Recently, Braun et al reported the heat capacity and low thermal conductivity of MgO-HEOs.[15] Meisenheimer et al reported antiferromagnetic order in MgO-HEO by studying the exchange anisotropy of permalloy/MgO-HEO heterostructures.[16] However, up to date, there are no reports on the magnetic structure on bulk samples.

The observation of new functionality in MgO-HEO triggered the search for other types of high entropy oxides beyond the rocksalt structure. Fluorite,[17-18] perovskite[4-6] and spinel[7] high entropy oxides have been reported. Besides structural studies, the fluorite high entropy oxides have been reported to have very low thermal conductivity,[17] making these materials promising candidates for thermal insulators and satisfying one of the criteria for high performance thermoelectric materials. More recently, Witte et al studied the magnetic properties of perovskite high entropy oxides and found predominant antiferromagnetic behavior with a small ferromagnetic contribution using magnetometry and Mössbauer spectroscopy.[19]

In this communication, we report the magnetic structure of the rocksalt high entropy oxide using neutron powder diffraction. This is the first report of magnetic structure for high entropy oxides.

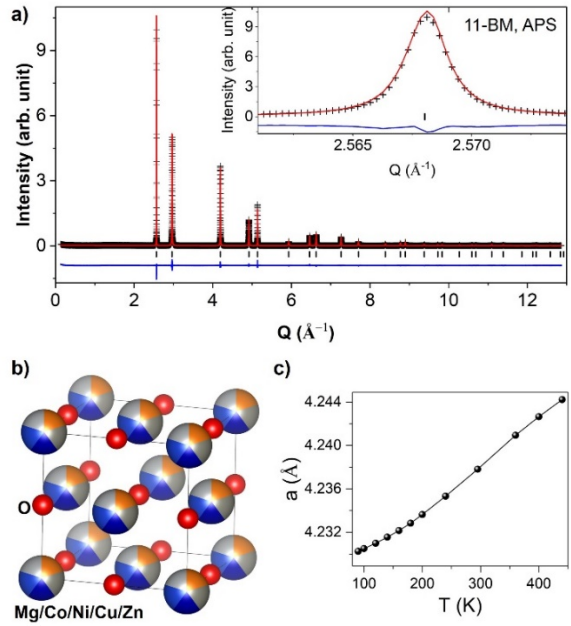

**Figure 1**. Crystal structure of $(Mg_{0.2}Co_{0.2}Ni_{0.2}Cu_{0.2}Zn_{0.2})O$. (a) Rietveld refinement of diffraction data as a function of $Q$ ($Q=4\pi sin\theta/\lambda$, $\lambda= 0.412758$ Å) collected at room temperature using synchrotron x-rays, (b) Crystal structure, (c) Temperature dependence of lattice parameters between 90 and 440 K.

MgO-HEO polycrystalline powders were synthesized using standard solid-state reaction techniques according to Rost et al.[3] Chemical analysis was carried out using a Bruker Quantax 70 energy dispersive spectroscopy (EDS) system on a Hitachi TM-3000 scanning electron microscope. Mg, Co, Ni, Cu and Zn were found to distribute homogenous at micron level (see Figure S1). EDS measurements confirm that the ratio of Mg, Co, Ni, Cu and Zn is essentially 1:1:1:1:1 (see Figure S2).

We investigated the average crystal structure of MgO-HEO using high resolution synchrotron x-ray powder diffraction at 11-BM at the Advanced Photon Source, Argonne National Laboratory. The relative intensities of the Bragg peaks (111) and (200) deviates from the ideal rocksalt lattice, indicating anisotropic strain broadening. Such a phenomenon has been observed by Berardan et al.[11] The Rietveld refinement on the data collected at room temperature with a single phase of $Fm$-$3m$ converged to $R_{wp}$=12.5% (Figure 1a, see Supporting Information for details). A sketch of this MgO-HEO structure is shown in Figure 1b. In such a structure, Mg, Co, Ni, Cu, and Zn reside on the same site (000, Wyckoff position 4$a$) with octahedral oxygen coordination, and oxygen locates at (½,½,½) (Wyckoff position 4$b$). No structural phase transition was observed within our resolution down to 90 K, as no extra peaks appear or peak splitting occurs on cooling (see Figure S3). The temperature dependence of lattice parameter from the Rietveld refinements is shown in Figure 1c, with *a* shrinking upon cooling. Linear thermal expansion coefficient is estimated to be $11\times10^{-6}$ K$^{-1}$ at room temperature, which is comparable to that of MgO.

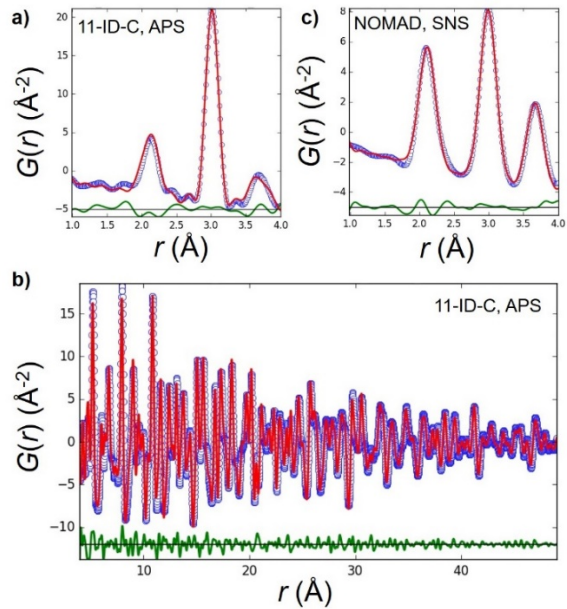

**Figure 2**. Local structure of $(Mg_{0.2}Co_{0.2}Ni_{0.2}Cu_{0.2}Zn_{0.2})O$. (a) x-ray PDF in the 1-4 Å range. (b) x-ray PDF in the 1-49 Å range. (c) neutron PDF in the 1-4 Å range. Empty blue circles: experimental data; red line: fitted structure; green line: residual. Residuals have been shifted by -5 for (a, c) and -12 for (b).

The local structure of MgO-HEO obtained from total scattering of synchrotron x-rays and neutrons, which yield better contrast for Co, Ni, Cu and Zn, can be described using the rocksalt structure with random cationic distribution. The x-ray pair distribution function (PDF) fit to the *Fm*-3*m* structure in the range of 1-49 Å (Figure 2b) converged to $R_w$=15.6%, demonstrating long range order of the atoms, consistent with x-ray powder diffraction analysis. The x-ray PDF fit in the 1-4 Å range converged to $R_w$=8.9% (Figure 2a), indicating that also the local structure can be described using *Fm*-3*m*. The neutron PDF in the 1-4 Å range was also fitted well using the cubic structure ($R_w$=7.6%, see Figure 2c), confirming similar local environments of Co, Ni, Cu and Zn. The shape of the nearest-neighbor peak is somewhat asymmetric, deviating from a single Gaussian function with extra intensity on the high-*r* side. This indicates that the local structure is not ideal rock-salt, which is consistent with a slightly distorted local environment resulting from Jahn-Teller effect of $Cu^{2+}$.[11-14] However, such an effect cannot be in a static, cooperative manner.

In order to characterize the oxidation states of cations, we carried out x-ray absorption spectroscopy on MgO-HEO. The oxidation states of Ni, Co, and Cu are 2+, as evidenced by the *L* edge x-ray absorption of Ni, Co and Cu as a function energy (see Figure S4). Our result is in good agreement with Rost's EXAFS report.[12]

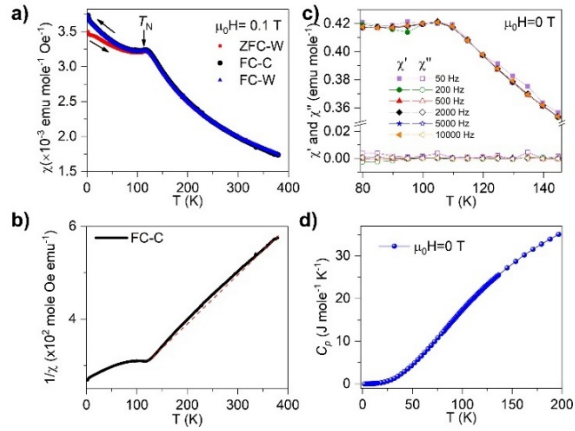

**Figure 3**. Physical properties of $(Mg_{0.2}Co_{0.2}Ni_{0.2}Cu_{0.2}Zn_{0.2})O$. (a) DC magnetic susceptibility measured with a magnetic field of 0.1 T. ZFC-W, FC-C and FC-W are short for zero-field cooling with data collection on warming, field cooling with data collection on cooling, and field-cooling with data collection on warming, respectively. (b) Inverse magnetic susceptibility as a function of temperature. The dashed line is only a guide for the eye. (c) Real and imaginary part of the AC magnetic susceptibility measured without external magnetic field at various frequencies. (d) Heat capacity measured under a magnetic field of 0 T.

We now present the physical properties of MgO-HEO, including magnetic susceptibility, heat capacity and thermal conductivity. Figure 3a shows the DC magnetic susceptibility as a function of temperature measured with a magnetic field of 0.1 T on warming after zero field cooling (ZFC-W), cooling with magnetic field (FC-C) and warming with field (FC-W). An anomaly is immediately identified, which suggests an antiferromagnetic order. The transition temperature is determined to be 113 K by the peak position of $\partial(\chi T)/\partial T$. The magnetic susceptibility continues to increase below 113 K. This suggests that not all of the local moments participate in the long range order, which is somewhat expected for such a strongly disordered system. The inverse magnetic susceptibility, $1/\chi$, is not linear above 113 K up to 380 K (see Figure 3b), suggesting non-Curie-Weiss behavior (see a modified Curie-Weiss fit in Figure S5). Such a deviation from the Curie-Weiss law may originate from short-range fluctuations, which is corroborated by the absence of an anomaly in heat capacity, the nonzero intensity of magnetic peaks above 113 K in neutron diffraction, and magnetic excitations at room temperature in inelastic neutron scattering, as will be discussed below. Magnetization was also measured as a function of magnetic field at 300 K and 2 K, and both show linear behavior (See Figure S6), which is consistent with paramagnetism at room temperature and an antiferromagnetic or spin glass state at low temperature. The real and imaginary part of the AC magnetic susceptibility as a function of temperature measured at various frequencies without magnetic field are presented in Figure 3c. No frequency dependence of the peak at 113 K is seen, and no peak in the imaginary part occurs near the transition. These observations are consistent with a long-range antiferromagnetic order, and inconsistent with spin-glass freezing. Figure 3d shows the heat capacity as a function of temperature without applied magnetic field. Surprisingly, no anomaly is observed. One plausible explanation is that the antiferromagnetic transition is not sharp and the entropy release occurs in a broad temperature range, corroborated by short range magnetic fluctuations above Néel temperature and magnetic excitations at room temperature (see Figure 4d and Figure S8). The thermal conductivity of MgO-HEO exhibits phonon glass behavior (see Figure S7), which is expected as it contains extreme chemical disorder.

To understand the nature of the phase transition at 113 K observed in DC magnetic susceptibility, we performed neutron powder diffraction at HB2A at HFIR, ORNL.[20-21] The nuclear structure was Rietveld refined from data collected at 160 K using the cubic structure $Fm$-$3m$ with the fit converging to $R_{wp}$=12.3%. Four extra peaks at $Q$=1.277 Å$^{-1}$, 2.455 Å$^{-1}$, 3.233 Å$^{-1}$ and 3.856 Å$^{-1}$ that do not coincide with nuclear Bragg reflections are observed at 2 K (see Figure 4a). The absence of these peaks at 160 K and the lack of a structural transition in synchrotron x-ray diffraction data confirm that they arise from antiferromagnetic order. Indexing of these peaks leads to a commensurate propagation vector $k$=(½, ½, ½). The magnetic structure was determined based on representation analysis using SARAh[22-23] and FullProf.[24] For the space group $Fm$-$3m$ with propagation vector (½, ½, ½) and magnetic ions on site $4a$ (0,0,0), there are two irreducible representations (IRs) ($\Gamma_3$ and $\Gamma_5$), following the labelling scheme of Kovalev,[25] see Supporting Information for basis vector components and Table S1. Both IRs were tested against the data and only $\Gamma_5$ found to fit all the observed reflections (see Figure 4b). The magnetic structure consists of ferromagnetic sheets in the (111) planes, with spins antiparallel between two neighboring planes, as shown in Figure 4c. The ordered magnetic moment is 1.4(1) $\mu_B$, which is somewhat smaller than 2 $\mu_B$ expected for the average pure spin moment calculating from 1/3 of 3 $\mu_B$ from Co$^{2+}$ ($S$=3/2), 1/3 of 2 $\mu_B$ from Ni$^{2+}$ ($S$=2/2), and 1/3 of 1 $\mu_B$ from Cu$^{2+}$ ($S$=1/2). The observation of reduced moment from experiments is understandable due to suppression of ordering throughout the lattice because of extreme chemical disorder and a significant amount of nonmagnetic ions, covalency (some of the moment is spread out over the oxygen), or magnetic fluctuations. Figure 4d shows the temperature dependence of the intensity at $Q=1.277$ Å$^{-1}$ through the cooling and warming processes. Notably, intensity above background is observed above $T_N$ determined from magnetic susceptibility. A plausible explanation for this observation is that neutron diffraction measurements cover a certain inelastic energy range around $E=0$ meV, thus suggesting dynamic fluctuations. Such an observation strongly suggests that the magnetic order occurs in a relatively wide temperature range, in good agreement with the absence of an anomaly in heat capacity.

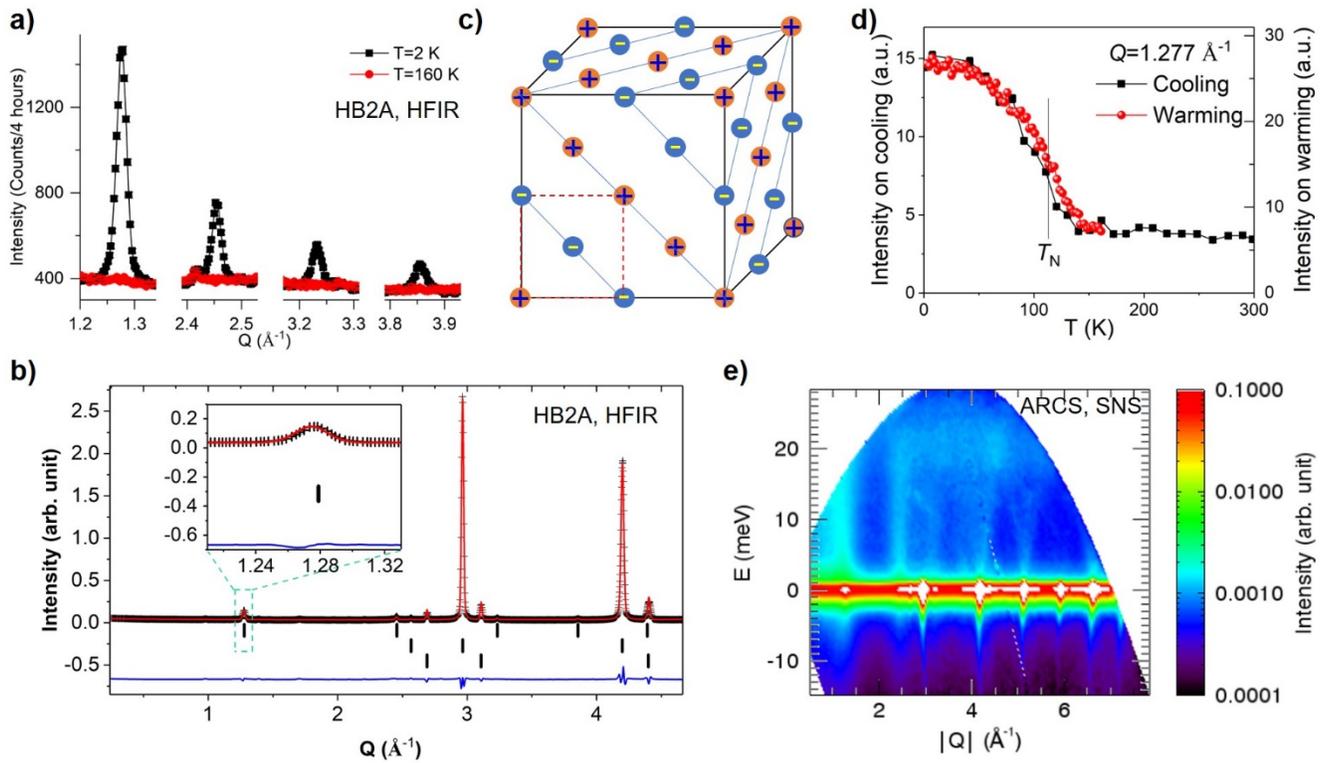

*Figure 4*. Magnetic structure and excitations of (Mg$_{0.2}$Co$_{0.2}$Ni$_{0.2}$Cu$_{0.2}$Zn$_{0.2}$)O. (a) Neutron diffraction peaks at selected $Q$ ($Q=4\pi\sin\theta/\lambda$, $\lambda= 2.41$ Å) ranges at 2 K and 160 K. (b) Rietveld refinement for the 2 K data with fit quality $R_{wp}=11.7\%$. Top: magnetic; Middle: nuclear; Bottom: Al can. Inset shows the magnification for the 1.21-1.33 Å$^{-1}$ range. (c) Scheme of magnetic structure. +: spin up; -: spin down. (d) Temperature dependence of the intensities at $Q=1.277$ Å$^{-1}$ on cooling and warming processes. (e) Dynamic structure factor at 100 K with $E_i=30$ meV.

The antiferromagnetic origin of the phase transition at 113 K is also evidenced from collective spin dynamics, *i.e.* magnons, which were measured on polycrystalline powders by selecting two incident energy of neutrons ($E_i=140$ and 30 meV) at the Wide Angular-Range Chopper Spectrometer (ARCS) spectrometer at the Spallation Neutron Source (SNS). Figure 4e presents the dynamic structure factor $S(Q,E)$ for the rocksalt MgO-HEO at 100 K with $E_i$ of 30 meV. Strong magnons are observed at $Q=1.277$ Å$^{-1}$ with a steep slope, and die out quickly with the increasing of $Q$. These magnons survive up to 300 K (see Figure S8), indicating strong interaction between magnetic ions even well above $2 \times T_N$.

The existence of antiferromagnetic order in the rocksalt high entropy oxide is not unexpected. Both CoO and NiO have the NaCl-type crystal structure, and are antiferromagnetic with propagation wavevector of $k=(½, ½, ½)$ and Néel temperature of 291 K for CoO and 523 K for NiO.[26] The substitution with nonmagnetic ions such as Mg or Zn suppresses long range antiferromagnetic order in these FCC antiferromagnets, as was exemplified by the $Co_{1-x}Mg_xO$ solid solution which becomes a spin glass for $x>0.53$ and a paramagnet for $x>0.87$,[27] and by $Ni_{1-x}Mg_xO$ which is a frustrated antiferromagnet ($0.37<x\leq0.6$), spin glass ($0.6<x\leq0.75$) and paramagnet state ($x\geq0.8$).[28] Supposing that $x=0.4$ applies for the case of MgO-HEO, it would correspond to the antiferromagnetic state of $Co_{1-x}Mg_xO$ or the frustrated antiferromagnetic state of $Ni_{1-x}Mg_xO$.

However, there are several major differences between MgO-HEO and the randomly diluted FCC system. Firstly, the magnitude of the spin exchange interactions in MgO-HEO are expected to be different from diluted FCC systems due to the existence of $Cu^{2+}$ ($d^9$, $S=1/2$) on the 4$a$ site in addition to $Co^{2+}$ ($d^7$, $S=3/2$) and $Ni^{2+}$ ($d^8$, $S=2/2$). In the rocksalt structure, the transition metal (TM) ions are connected via oxygen atoms with TM-O-TM bond angle 180°. The magnetic exchange interactions are superexchange interactions. Based on Kannamori-Goodenough rules,[29-30] the TM-O-TM coupling in MgO-HEO can be effectively AFM or FM, depending on the electronic configuration of the two coupled TM ions. AFM interactions dominates for Co (or Ni, Cu)-O-Co(or Ni, Cu). For NiO and CoO based diluted systems, the Néel temperature is dominated by the next-nearest-neighbor exchange interaction, $J_2$.[31-32] For example, the spin exchange interactions in CoO are reported to be $J_2= -25.54$ meV and $J_1= 8.00$ meV (nearest-neighbor interaction).[33] The exchange interactions in MgO-HEO are the key to understand its long range antiferromagnetism. To address this, theoretical calculations on spin exchange interactions are required. Secondly, no structural transition or deformation has been observed across the magnetic transition in MgO-HEO, in sharp contrast to the diluted FCC systems with nonmagnetic ions, which are tetragonal, rhombohedral or monoclinic in the antiferromagnetic state.[34-37] Thirdly, the absence of a λ-shape peak around the magnetic transition in heat capacity in MgO-HEO contrasts with that in CoO.[38] Such characteristics are related to the large degree of chemical disorder stabilized by high configuration entropy. It will be very interesting to study the magnetic transitions and magnetic structures of other types of high entropy oxides to see if these features are general. Along this line, the perovskite $R(Cr_{0.2}Mn_{0.2}Fe_{0.2}Co_{0.2}Ni_{0.2})O_3$ ($R$=La, Gd, Nd, Sm, Y)[19] are obvious candidates.

In summary, the rocksalt high entropy oxide is long range antiferromagnetically ordered below $T_N$=113 K, as evidenced by a cusp in DC magnetic susceptibility, magnetic peaks in neutron powder diffraction, and strong magnetic excitations in inelastic neutron scattering. The magnetic structure consists of ferromagnetic sheets in the (111) planes with spins antiparallel between two neighboring planes, similar to NiO or CoO. The ordered magnetic moment is 1.4(1) $\mu_B$. This work opens a new chapter for understanding the magnetic properties of high entropy oxides, a fascinating class of materials stabilized by high configuration entropy.

**ACKNOWLEDGMENT**

This material is based upon work supported by the U.S. Department of Energy, Office of Science, Office of Basic Energy Sciences, Materials Sciences and Engineering Division. The use of Oak Ridge National Laboratory's Spallation Neutron Source and High Flux Isotope Reactor was supported by the Scientific User Facilities Division, Office of Basic Energy Sciences, U.S. Department of Energy. Use of the Advanced Photon Source, an Office of Science User Facility operated for the US Department of Energy (DOE) Office of Science by Argonne National Laboratory, was supported by the US DOE under Contract DE-AC02-06CH11357. J.Z. would like to thank Mr. Liang Wang for his help with x-ray pair distribution function data at 11-ID-C. J.Z. acknowledges the fourth "Modern Methods in Rietveld Refinement for Structural Analysis" workshop held at the Advanced Photon Source of Argonne National Laboratory in close partnership with Bruker-AXS, ANL, and the National Science Foundation. The authors thank Drs. M. E. Manley, V. R. Cooper, T. Z. Ward, Y. Sharma, K. Pitike, and Mr. A. R. Mazza for fruitful discussions.